\documentclass[twocolumn,floatfix,showkeys,nofootinbib,reprint]{revtex4-1}
\usepackage{epsfig}
\usepackage{latexsym}
\usepackage{xspace}
\usepackage[colorlinks=true,linktocpage=true,linkcolor=blue,citecolor=blue,allcolors=blue]{hyperref}
\usepackage[utf8]{inputenc}
\usepackage{indentfirst}
\usepackage{enumerate}
\usepackage{listings}
\usepackage{color}

\usepackage[caption=false,position=top]{subfig}

\usepackage{amsmath}
\usepackage{amssymb}
\usepackage[english]{babel}
\usepackage{url}

\newcommand{\sNN}{\sqrt{s_{\rm NN}}}

\newcommand{\bvar}{\mathbf}

\newcommand{\eq}[1]{\begin{align} #1 \end{align}}
\newcommand{\mean}[1]{\langle #1 \rangle}

\begin{document}

\title{Magnetic field effect on hadron yield ratios and fluctuations in a hadron resonance gas}

\author{Volodymyr~Vovchenko}
\affiliation{Physics Department, University of Houston, 3507 Cullen Blvd, Houston, TX 77204, USA}

\begin{abstract}
This work studies the influence of an external magnetic field on hadron yields and fluctuations in a hadron resonance gas by performing calculations within an updated version of the open-source Thermal-FIST package.
The presence of the magnetic field has a sizable influence on certain hadron yield ratios. Most notably, it leads to enhanced $p/\pi$ and suppressed $n/p$ ratios, which may serve as a magnetometer in heavy-ion collisions.
By attributing the centrality dependence of the $p/\pi$ ratio in Pb-Pb collisions at 5.02~TeV measured by the ALICE Collaboration entirely to the magnetic field, its maximal strength at freeze-out is estimated to be $eB \simeq 0.12$~GeV$^2 \simeq 6.3 m_\pi^2$ in peripheral collisions.
The magnetic field also enhances various conserved charge susceptibilities, which is qualitatively consistent with recent lattice QCD data and is driven in the HRG model by the increase of hadron densities.
However, the variances of hadrons do not show any enhancement when normalized by the means, therefore, measurements of second-order fluctuations in heavy-ion collisions appear to offer limited additional information about the magnetic field relative to mean multiplicities.
\end{abstract}



\maketitle

\section{Introduction}

The properties of QCD matter can change under the presence of an external magnetic field $eB$~\cite{Andersen:2014xxa,Miransky:2015ava}, which has attracted considerable interest recently~\cite{Mueller:2015fka,Finazzo:2016mhm,Farias:2016gmy,Singh:2017nfa,Bali:2017ian,Ding:2020hxw,Xu:2020yag}.
Strong magnetic field effects can be found in different environments, including early universe~\cite{Vachaspati:1991nm}, neutron stars~\cite{Harding:2006qn}, and, in particular, in non-central collisions of relativistic heavy ions~\cite{Kharzeev:2007jp,Skokov:2009qp,Huang:2015oca}, where they can leave observables imprints such as chiral magnetic effect~\cite{Fukushima:2008xe}.
The effect of magnetic field of the chiral QCD transition at vanishing baryon densities can be studied through first-principle lattice QCD simulations~\cite{Bali:2011qj,Bali:2012zg} indicating inverse magnetic catalysis of the transition temperature and a possibility of a critical point in temperature-magnetic field plane~\cite{Endrodi:2015oba,DElia:2021yvk}.
Recent lattice simulations evaluate the effect of magnetic field on conserved charge susceptibilities~\cite{Ding:2021cwv,Ding:2023bft}, which serve as important probes of QCD degrees of freedom~\cite{Koch:2005vg}.

The hadronic part of the QCD phase diagram is commonly described in the framework of the hadron resonance gas~(HRG) model~\cite{Hagedorn:1965st,Karsch:2003vd}, which has long history of applications to describe hadron yields in heavy-ion collisions and extract chemical freeze-out parameters~\cite{Cleymans:1992zc,Becattini:2005xt,Andronic:2017pug}.
These analyses normally employ the standard HRG model without the effects of an external magnetic field.
The HRG model in an external magnetic field has been formulated in Ref.~\cite{Endrodi:2013cs}, where a magnetized hadron spectrum was introduced, and the influence of the magnetic field on the thermodynamic properties of the HRG model was studied.
The HRG model in an external magnetic field has later been used to study the behavior of conserved charge susceptibilities~\cite{Bhattacharyya:2015pra,Ding:2021cwv}, properties of a non-ideal HRG~\cite{Sahoo:2023vkw}, and the magnetic field effect on chemical freeze-out curve and net-charge fluctuations~\cite{Fukushima:2016vix}.

Recently, conserved charge susceptibilities have been evaluated at finite temperature and magnetic field in lattice QCD~\cite{Ding:2023bft}.
Based on its strong sensitivity to the magnetic field, a baryon-charge correlator $\chi_{11}^{BQ}$ has been suggested as a possible magnetometer in heavy-ion collisions.
The present work explores in detail the impact of magnetic field on hadron abundances in the HRG model and emphasizes that the enhancement of conserved charge susceptibilities in the HRG model is entirely driven by the enhancement of the hadron densities.
In particular, the strong enhancement of the doubly-charged spin-3/2 $\Delta^{++}$ yields at non-zero $eB$ leaves an impact on the final proton yields due to feeddown.
It is argued that centrality dependence of hadron ratios such as $p/\pi$ and, potentially, $n/p$ provide a cleaner probe of a sizable magnetic field at freeze-out in heavy-ion collisions, if one exists there, compared to fluctuations.

The paper is structured as follows.
In Sec.~\ref{sec:HRG} the formulation of the HRG model in external magnetic field reviewed and its implementation in the Thermal-FIST package is described.
The influence of magnetic field on various hadron yields and their ratios measurable in heavy-ion collisions is discussed in Sec.~\ref{sec:yields}.
Section~\ref{sec:fluctuations} presents the behavior of fluctuations and comparison with corresponding quantities from lattice QCD.
Discussion and summary in Sec.~\ref{sec:summary} close the article.

\section{HRG model in external magnetic field}
\label{sec:HRG}

The HRG model in an external magnetic field $eB$ has been formulated in Ref.~\cite{Endrodi:2013cs}, and this formulation is utilized here. The presence of magnetic field leads to several modifications relative to the standard HRG model: (i) vacuum contribution; (ii) modification of the energy levels; and (iii) discretization of transverse momenta.

The vacuum contributions consist of pure magnetic energy part as well as that from the interaction of the magnetic field with virtual hadrons~\cite{Endrodi:2013cs}. This contribution is independent of the chemical potentials and does not affect hadron abundances or susceptibilities.
It is thus omitted in the present work.

The energy levels and momentum states of neutral particles are assumed to be unaffected by the magnetic field, therefore, $E_n(\bvar p) = \sqrt{\bvar p^2 + m^2}$ holds for neutral particles.
For charged particles, their transverse momenta become discretized and the corresponding states are enumerated by Landau levels, $l = 0, 1, \ldots$.
Assuming a uniform magnetic field along $z$-axis, the energy levels are written as a function of $p_z$, $l$, and the spin projection $s_z$,
\eq{\label{eq:Ec}
E_c(p_z, l, s_z) = \sqrt{p_z^2 + m^2 + 2 |q| B (l + 1/2 - s_z)}.
}
Here the gyromagnetic ratio of $g = 2|q/e|$ is taken.
In the absence of detailed experimental knowledge of gyromagnetic ratios for different hadrons, $g = 2|q/e|$ is taken here to hold for all charged hadrons, as motivated in Refs.~\cite{Ferrara:1992yc,Endrodi:2013cs}.
The momentum integrals are modified as
\eq{
(2 s + 1) \, \int \frac{d^3 \bvar p}{(2\pi)^3} \rightarrow \frac{|q|B}{2\pi} \sum_{s_z = -s}^s \sum_{l=0}^{\infty} \int \frac{d p_z}{2\pi}
}

The thermal part of the pressure contains contributions from all neutral and charged particles
\eq{
\label{eq:p}
p = \sum_{i \in neutral} p_n^i + \sum_{i \in charged} p_c^i,
}
where
\eq{\label{eq:pn}
p_n^i & = \eta_i \frac{(2s_i + 1)T}{2 \pi^2} \int_0^\infty dp \, p^2 \, \ln\left[ 1 + \eta_i e^{-(E_n - \mu_i)/T} \right] \nonumber \\
& = \frac{(2s_i + 1)}{6 \pi^2} \int_0^\infty dp \, \frac{p^4}{E_n} \left[\exp\left(\frac{E_n - \mu_i}{T}\right) + \eta_i\right]^{-1} \nonumber \\
& = \frac{(2s_i + 1) m_i^2 T^2}{2 \pi^2} \sum_{k=1}^{\infty} (-\eta_i)^{k+1} \frac{e^{k\mu_i/T}}{k^2} K_2(km_i/T)
}
and
\eq{\label{eq:pc}
p_c^i & = \eta_i \frac{|q_i| B T}{2 \pi^2} \sum_{s_z = -s}^s \sum_{l=0}^{\infty}\int_{-\infty}^{\infty} dp_z \, \ln\left[ 1 + \eta_i e^{-(E_c - \mu_i)/T} \right] \nonumber \\
& = \frac{|q_i| B}{2 \pi^2} \sum_{s_z = -s}^s \sum_{l=0}^{\infty}\int_{-\infty}^{\infty} dp_z \, \frac{p_z^2}{E_c} \left[\exp\left(\frac{E_c - \mu_i}{T}\right) + \eta_i\right]^{-1} \nonumber \\
& = \frac{|q_i| B T}{2 \pi^2} \sum_{s_z = -s}^s \sum_{l=0}^{\infty} \varepsilon_c \sum_{k=1}^{\infty} (-\eta_i)^{k+1} \frac{e^{k\mu_i/T}}{k} K_1(k\varepsilon_c/T)
}
Here $\varepsilon_c = \sqrt{m^2 + 2 |q| B (l + 1/2 - s_z)}$, $s_z$ is the spin of hadron species $i$, $m_i$ is its mass, $\mu_i = b_i \mu_B + q_i \mu_Q + s_i \mu_S$ is its chemical potential, and $\eta_i = \pm 1$ corresponds to Fermi~(Bose) statistics.
Each line in Eqs.~\eqref{eq:pn} and \eqref{eq:pc} corresponds to an equivalent expression for the partial pressure.

Note that the pressure defined by Eqs.~\eqref{eq:p}-\eqref{eq:pc} is isotropic and corresponds to the derivative of the grand potential with respect to the volume at constant magnetic field.
If, instead, the flux $\Phi$ of the magnetic field is kept constant, such as the case in lattice QCD simulations~\cite{Bali:2013esa}, the pressure is anisotropic.
The longitudinal pressure $p_z$ is still defined by Eqs.~\eqref{eq:p}-\eqref{eq:pc} while the transverse pressure is smaller, $p_{x,y} = p_z - eB \mathcal{M}$, where $\mathcal{M} = (\partial p_z/\partial B)_{T,\mu}$ is the magnetization.
It has been argued that the choice of scheme does not affect observables such as energy density~\cite{Bali:2013esa} and, presumably, susceptibilities, but it remains to be seen if this is the case for hadron densities under conditions encountered in heavy-ion collisions. 
Here, the hadron number densities are
evaluated in the standard way, as derivatives of the (longitudinal) pressure with respect to $\mu_i$:
\eq{\label{eq:nn}
n_n^i & = \frac{(2s_i + 1)}{2 \pi^2} \int_0^\infty dp \, p^2 \left[\exp\left(\frac{E_n - \mu_i}{T}\right) + \eta_i\right]^{-1} \nonumber \\
& = \frac{(2s_i + 1) m_i^2 T}{2 \pi^2} \sum_{k=1}^{\infty} (-\eta_i)^{k+1} \frac{e^{k\mu_i/T}}{k} K_2(km_i/T),
}
and
\eq{\label{eq:nc}
n_c^i & = \frac{|q_i| B}{2 \pi^2} \sum_{s_z = -s}^s \sum_{l=0}^{\infty}\int_{-\infty}^{\infty} dp_z \left[\exp\left(\frac{E_c - \mu_i}{T}\right) + \eta_i\right]^{-1} \nonumber \\
& = \frac{|q_i| B}{2 \pi^2} \sum_{s_z = -s}^s \sum_{l=0}^{\infty} \varepsilon_c  \sum_{k=1}^{\infty} (-\eta_i)^{k+1} e^{k\mu_i/T} K_1(k\varepsilon_c/T).
}

Higher-order derivatives of the pressure with respect to chemical potentials define the various susceptibilities. For the conserved charges, these read
\eq{
\chi^{BQS}_{lmn} = \frac{\partial^{l+m+n}(p/T^4)}{\partial(\mu_{B}/T)^{l} \, \partial(\mu_{Q}/T)^{m} \, \partial(\mu_{S}/T)^{n}},
}
where partial derivatives are taken at fixed values of the temperature and other chemical potentials.
In the ideal HRG model, the hadrons numbers are uncorrelated, therefore, conserved charge susceptibilities can be expressed in terms of individual hadron number susceptibilities
\eq{\label{eq:chihadr}
\chi^{BQS}_{lmn} = \sum_{i} b_i^l \, q_i^m \, s_i^n \, \frac{\partial^{l+m+n} (p_{n/c}^i / T^4)}{\partial (\mu_i/T)^{l+m+n}}~.
}
The partial derivatives in the right-hand-side of Eq.~\eqref{eq:chihadr} are taken at fixed temperature $T$.

The effect of the magnetic field on HRG thermodynamics at finite temperature has now been implemented in the open-source Thermal-FIST package~\cite{Vovchenko:2019pjl} since version 1.5~\cite{FIST1p5}. 
This is done through the modification of thermal integrals described above. 
This also allows us to combine the magnetic field effect with other HRG model extensions available in Thermal-FIST, including resonance decay feeddown, finite resonance widths~\cite{Vovchenko:2018fmh}, and van der Waals interactions~\cite{Vovchenko:2016rkn}.

The present implementation does not incorporate modifications of hadron masses in the presence of the magnetic field.
This effect has been studied both in effective QCD models~\cite{Bandyopadhyay:2016cpf,Ghosh:2016evc,Avancini:2016fgq,Coppola:2019uyr,Ayala:2020muk,Coppola:2023mmq} and lattice QCD~\cite{Bali:2017ian,Endrodi:2019whh,Ding:2022tqn}, in particular affecting neutral hadrons whose masses tend to become smaller.
Although a consistent description would likely require incorporating interactions beyond the standard HRG picture, this effect could also be included through a phenomenological modification of hadron masses. 
This modification is left for future work.

One should also note that decay rates are expected to be modified in the presence of a magnetic field, although it remains a complicated task for calculating these modifications~\cite{Jaber-Urquiza:2023sal}.
These may modify both the primordial resonance abundances, especially if finite widths are taken into account, as well as their decay feeddown.
In the context of heavy-ion collisions, one should note that decay feeddown does take place at a later stage compared to the chemical freeze-out of primordial abundances. 
Effects of the magnetic may well diminish by that stage, making the use of vacuum branching ratios an appropriate choice.
In all subsequent calculations presented here, the vacuum branching ratios are used to calculate decay feeddown.

One should mention one subtlety with regard to incorporating finite resonance widths and finite magnetic field effects simultaneously.
At sufficiently strong magnetic field, the critical mass $m_{\rm cr} = \sqrt{-2 |q| B (1/2 - s_z)}$ for $s_z < -1/2$ hadrons below which the energy levels in Eq.~\eqref{eq:Ec} become complex may become larger than the decay threshold.
In such a case, the energy-dependent Breit-Wigner spectral function used in Thermal-FIST is modified to exclude the masses $m < m_{\rm cr}$, thus avoiding this instability.

\section{Magnetic field and hadron yield ratios}
\label{sec:yields}

\subsection{Effect on hadron yields}

\begin{figure*}[t]
    \includegraphics[width=.49\textwidth]{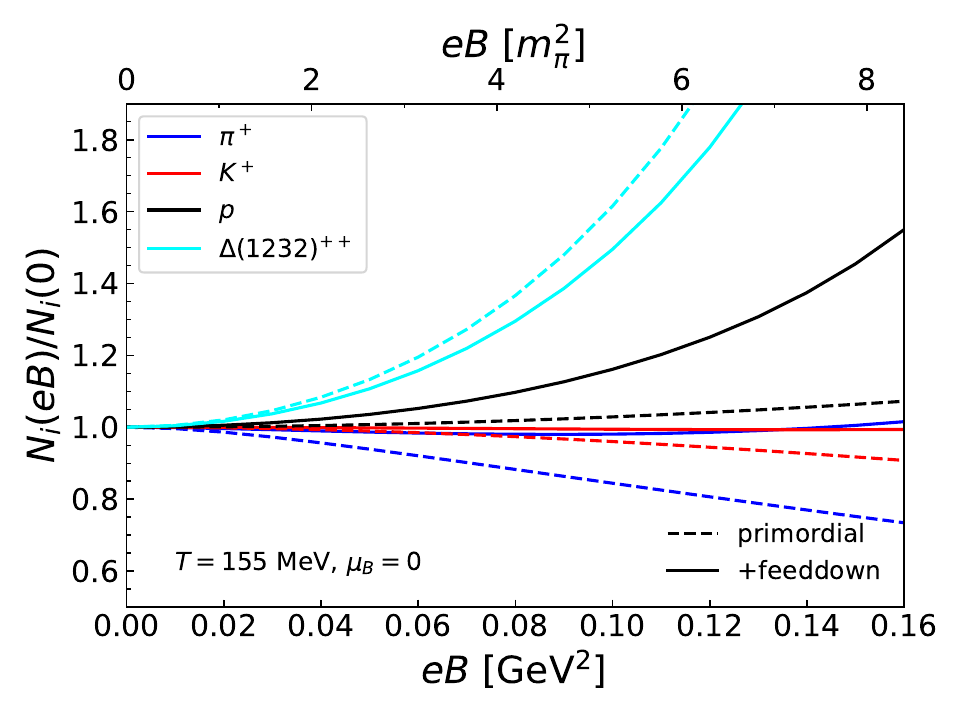}
    \includegraphics[width=.49\textwidth]{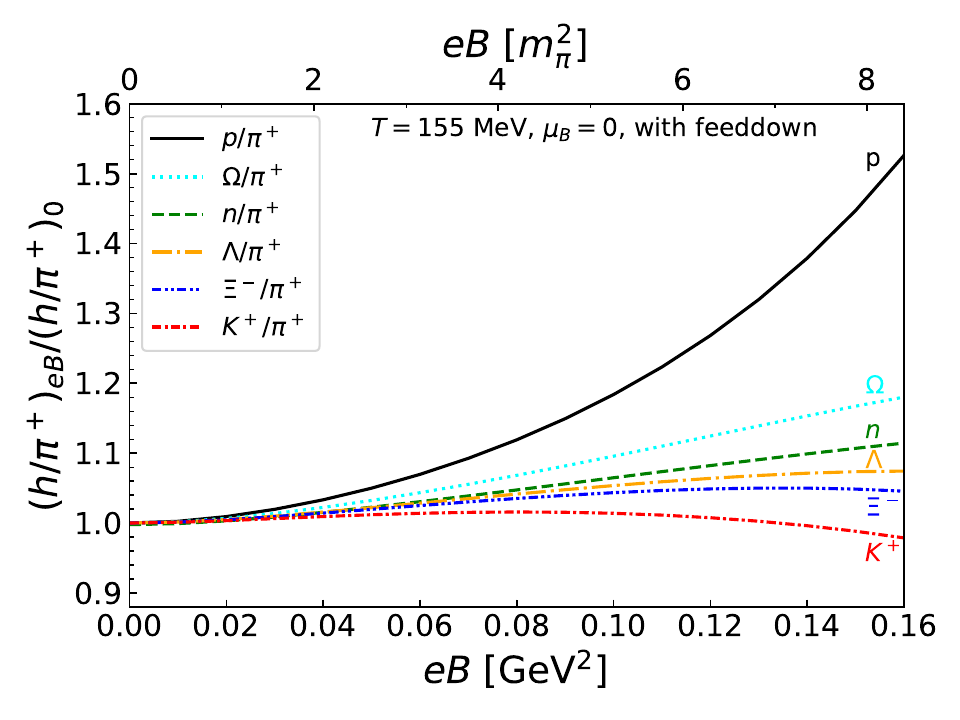}
    \caption{
    Dependence of various hadron yields~(left panel) and hadron-to-pion yield ratios~(right panel) on the external magnetic field $eB$ in HRG model at temperature $T = 155$~MeV and vanishing chemical potentials. All quantities are normalized by their corresponding values without magnetic field, $eB = 0$.
    The dashed lines in the left panel correspond to primordial hadron abundances, while the solid lines incorporate strong and electromagnetic decay feeddown.
    }
    \label{fig:yieldratios}
\end{figure*}

Magnetic field modifies hadron number densities of charged hadrons~[Eq.~\eqref{eq:nc}], especially those with large spin.
This can also be understood as charged hadrons acquiring effective mass due to the magnetic field, roughly as $m_{\rm eff} = \sqrt{m^2 + |q| B (1 - 2s_z)}$ as per Eq.~\eqref{eq:Ec}.
To illustrate this effect, in the left panel of Fig.~\ref{fig:yieldratios} the 
relative change in the densities of various hadrons due to the presence of magnetic field is shown.
The calculation is done for the temperature of $T = 155$~MeV and baryon chemical potential of $\mu_B = 0$, approximately corresponding to chemical freeze-out conditions in Pb-Pb collisions at the LHC~\cite{Andronic:2017pug}.
The dashed lines depict the relative change in primordial yields while the solid lines incorporate feeddown from strong and electromagnetic decays, the latter being appropriate for comparisons to experimental measurements in heavy-ion collisions.

One can see a strong enhancement of the $\Delta(1232)^{++}$ yield due to magnetic field, which can be attributed to it having spin $S = 3/2$ and being doubly charged, which reduce its effective mass considerably.
It should be noted, however, that applying the formalism to $S = 3/2$ particles such as $\Delta(1232)$ may be questionable, as the analysis of vacuum contributions to the pressure indicates that the presence of $S = 3/2$ (and above) states may make the system unstable~\cite{Endrodi:2013cs}.
It is also unclear how the decay properties of the resonances may be modified at non-zero $eB$.
On the other hand, the large effect of the magnetic field of $\Delta(1232)$ predicted by the present HRG model appears to be supported by lattice QCD data on baryon-charge susceptibility $\chi_{11}^{BQ}$, as discussed in the next section.
The primordial yield of $S = 1/2$ protons is not affected strongly by the magnetic field, but the total yield shows a sizable enhancement, reflecting significant feeddown from $\Delta$ resonances.
The primordial yields of $S = 0$ mesons, such as pions and kaons, are moderately suppressed by the magnetic field, but this effect is essentially washed out once feeddown is incorporated.

The right panel of Fig.~\ref{fig:yieldratios} shows the relative change due to the magnetic field of the various hadron-to-pion ratios that are commonly measured in the experiment. One can see the strongest effect in $p/\pi$ ratio, followed by $\Omega/\pi$ ratio, while the change in $\Lambda/\pi$ and $\Xi^-/\pi$ ratios is modest.
Thus, the $p/\pi$ ratio is expected to be the most sensitive to the possible presence of a non-zero magnetic field.
Interestingly, the effect on the $n/\pi$ ratio is considerably weaker compared to the $p/\pi$ ratio, reflecting the isospin symmetry-breaking nature of the magnetic field effect. 
In contrast to protons, primordial neutron yields are unaffected by the magnetic field, given that neutrons are neutral particles.
The final neutron yields are affected due to resonance decay feeddown, but, the effect is weaker than protons because neutrons do not contain feeddown from the doubly charged $\Delta(1232)^{++}$ resonance.
Sec.~\ref{sec:otherratios} discusses how the potential measurement of neutron yields, which is of course a very challenging task, could be utilized to probe the possible presence of magnetic field at freeze-out in heavy-ion collisions.

The above calculations neglect finite resonance widths.
This effect, however, can be important for hadron yields both a zero magnetic field~\cite{Vovchenko:2018fmh} and at finite $eB$.
To test the sensitivity of the results to finite widths the corresponding calculations have been performed here within the energy-dependent Breit-Wigner scheme~\cite{Vovchenko:2018fmh}.
The results show very similar modifications of the hadron-to-pion ratios due to non-zero $eB$ shown in Fig.~\ref{fig:eBLHCNchratios} with a slightly smaller magnitude.
As discussed before, the treatment of finite resonance widths at non-zero $eB$ does require extra care. However, the present analysis indicates that the inclusion of this effect is not expected to change the qualitative behavior.
The finite widths are omitted in the following.

\subsection{$p/\pi$ ratio as a magnetometer in heavy-ion collisions}

\begin{figure}[t]
    \includegraphics[width=.49\textwidth]{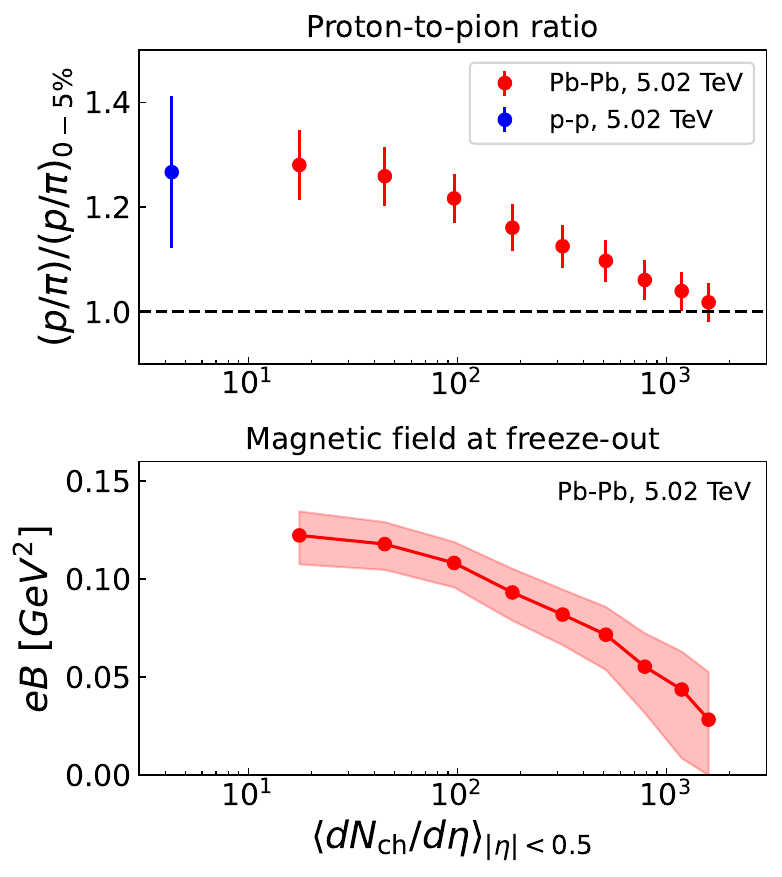}
    \caption{
    Upper panel: Multiplicity dependence of the $p/\pi$ ratio measured by the ALICE Collaboration~\cite{ALICE:2019hno} in Pb-Pb collisions at $\sNN = 5.02$~TeV. 
    The ratios are normalized by the $p/\pi$ ratio in $0-5\%$ collisions.
    Lower panel: Multiplicity dependence of the magnetic field $eB$ in hadron resonance gas model extracted from experimental data to match the enhancement of the $p/\pi$ ratio relative to central collisions. The band corresponds to the error propagation of the experimental data shown in the upper panel.
    }
    \label{fig:eBLHC}
\end{figure}

ALICE Collaboration has measured the $p/\pi$ ratio in Pb-Pb collisions as a function of centrality at $\sNN = 2.76$~TeV~\cite{Abelev:2013vea} and $\sNN = 5.02$~TeV~\cite{ALICE:2019hno}.
The experimental uncertainty for the measurements at $5.02$~TeV has been split into two contributions: (i) correlated and (ii) uncorrelated with centrality.
With the much smaller uncorrelated uncertainty, one can establish the suppression of the $p/\pi$ ratio in central collisions relative to non-central ones with sizeable statistical significance~(Fig.~\ref{fig:eBLHC}).
The suppression of $p/\pi$ ratio with centrality is commonly attributed to baryon annihilation in the hadronic phase, which is expected to be most relevant in central collisions~\cite{Becattini:2014hla,Vovchenko:2022xil}.

Here a different mechanism responsible for the centrality dependence of the $p/\pi$ ratio is explored. 
Namely, the larger $p/\pi$ ratio in non-central collisions relative to central ones is attributed to the presence of strong magnetic field.
As shown in the previous subsection, a non-zero magnetic field enhances the $p/\pi$ ratio, which is qualitatively consistent with the experimental measurement.
The strength of the magnetic field at chemical freeze-out in Pb-Pb collisions at the LHC needed to explain the experimental data is estimated here in the following way:
\begin{enumerate}
    \item It is assumed that the chemical freeze-out at all centralities occurs at $T_{\rm ch} = 155$~MeV, with hadron yields described by the HRG model in a non-zero magnetic field. 
    It is also assumed that the most central bin, $0-5\%$, corresponds to a vanishing magnetic field, $eB = 0$. The latter assumption holds on average due to symmetry, but it has been discussed that magnetic field can be large even in central collisions due to event-by-event fluctuations~\cite{Bzdak:2011yy,Deng:2012pc}. Taking into account this effect is a possible future refinement of the analysis.
    \item 
    The relative enhancement of the $p/\pi$ ratio at a given centrality over the $0-5\%$ centrality is mapped to the one given by non-zero magnetic field shown in Fig.~\ref{fig:yieldratios}.
    In this way one extracts the value of magnetic field $eB$ at a given centrality.
\end{enumerate}

The lower panel of Fig.~\ref{fig:eBLHC} depicts the multiplicity dependence of the magnetic field $eB$ extracted following the procedure described above. 
The magnetic field is strongest in peripheral collisions~(lowest multiplicities), reaching $eB \simeq (0.13 \pm 0.01)\text{GeV}^2 \simeq (6.6 \pm 0.6) m_\pi^2$.
Of course, the above analysis makes use of several simplifications, giving qualitative picture at best.
For instance, it is assumes that the magnetic field vanishes in central collisions due to symmetry. Although it holds on average, it has been discussed that magnetic field can be large even in central collisions due to event-by-event fluctuations~\cite{Bzdak:2011yy,Deng:2012pc}.
In addition, the magnetic field in the HRG model corresponds to its value in the local rest frame.
Due to the existence of radial flow and the fact that the magnetic field is not Lorenz-invariant, one can expect additional fluctuations in the values of the magnetic field in the HRG model even if $eB$ is constant in the collider frame.

Despite its qualitative nature, the above analysis is sufficient to indicate the necessary magnitude of the magnetic field at freeze-out for it to have a sizable effect on hadron yields, and possibly explain the centrality dependence of the $p/\pi$ ratio.
It is necessary to point out several limitations of the present analysis:

\begin{itemize}
    \item The extracted magnetic field of $eB \simeq (6.6 \pm 0.6) m_\pi^2$ is conceivable to exist in the early stage of the collision. It is, however, remarkably large for the freeze-out stage due to the expected fast temporal decay of the magnetic field, even in the presence of large electric conductivity and the strong medium response of QCD matter~\cite{Tuchin:2013ie}.
    A strong magnetic field at freeze-out should leave an imprint on other observables, such as inducing a difference between $\Lambda$ and $\bar{\Lambda}$ polarization. 
    However, no significant difference between $\Lambda$ and $\bar{\Lambda}$ polarization has been observed in Pb-Pb collisions at the LHC~\cite{ALICE:2019onw}, and an upper limit of $eB/m_\pi^2 \sim 0.044$ has been estimated for 5.02 TeV Pb-Pb collisions~\cite{ALICE:2022wpn}.

    \item The analysis assumes that the multiplicity dependence of $p/\pi$ ratio is driven entirely by the changing magnetic field. However, hadron yields can also be affected by the hadronic phase, which entails a prominent centrality-dependent effect~\cite{Becattini:2014hla}. 
    In particular, baryon annihilation suppresses the $p/\pi$ ratio in central collisions~\cite{Steinheimer:2012rd,Becattini:2012xb} and can plausibly describe the experimental data~\cite{Vovchenko:2022xil}.

    \item The $p/\pi$ ratio in $p-p$ collisions, where no strong magnetic field is expected, is consistent with measured $p/\pi$ value in peripheral Pb-Pb collisions and about 1.8$\sigma$ above the one in central collisions, see the blue point in Fig.~\ref{fig:eBLHC}.

    \item The freeze-out temperature was assumed to be $T_{\rm ch} = 155$~MeV independent of the magnetic field. It is feasible, however, that freeze-out temperature can change in the presence of the magnetic field. For instance, in Ref.~\cite{Fukushima:2016vix}, it has been argued that $T_{\rm ch}$ can become smaller as a possible manifestation of inverse magnetic catalysis, although sizable changes in $T_{\rm ch}$ may require larger magnetic fields than considered here.

\end{itemize}

For the above reasons, one can argue that the magnetic field is unlikely to be the primary driver behind the measured multiplicity dependence of the $ p/\pi$ ratio in Pb-Pb collisions at LHC. 
Nevertheless, in the following, the behavior of other hadron yield ratios is explored to test the possibility of the strong magnetic field at chemical freeze-out.

\subsection{Other hadron yield ratios}
\label{sec:otherratios}

\begin{figure}[t]
    \includegraphics[width=.49\textwidth]{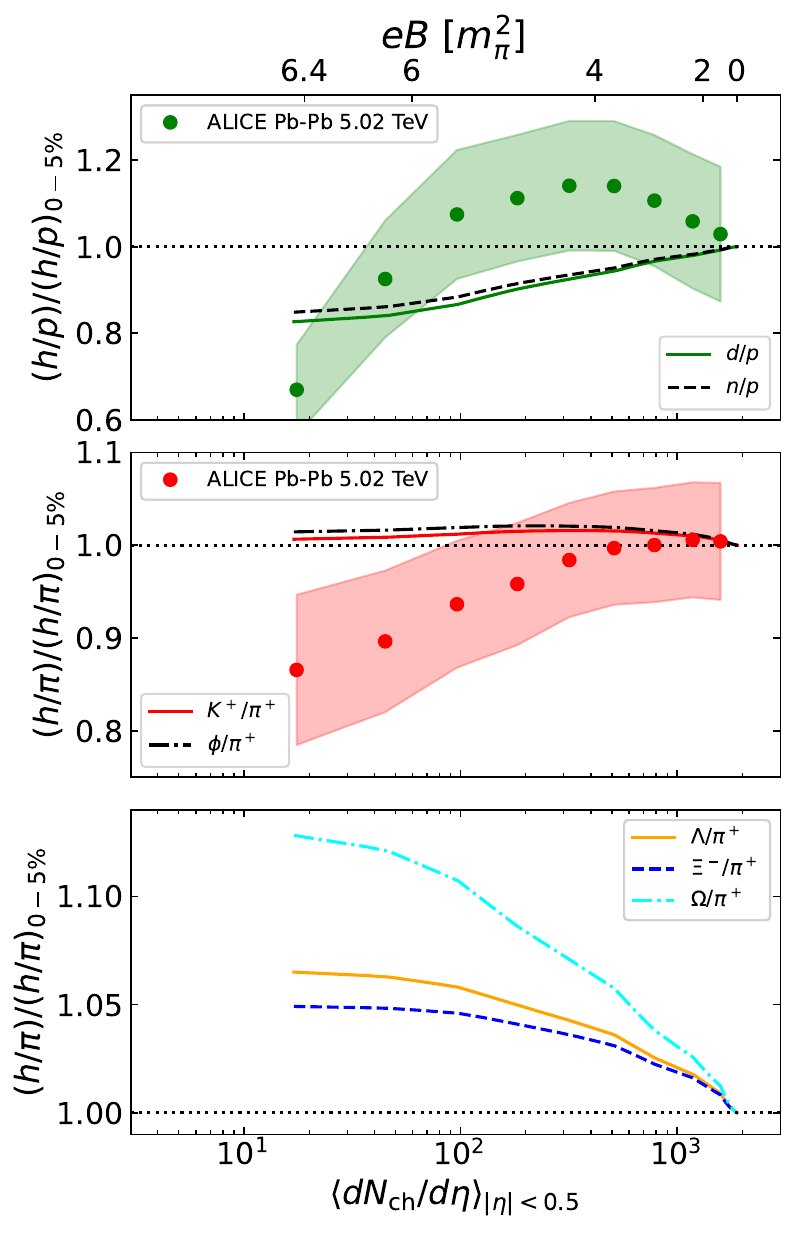}
    \caption{
    Multiplicity dependence of hadron yield ratios $d/p$ and $n/p$~(upper panel), $K^+/\pi^+$, $\phi/\pi^+$~(middle panel) and $\Lambda/\pi^+$, $\Xi^-/\pi^+$, and $\Omega/\pi^+$~(lower panel) in Pb-Pb collisions at  LHC energies calculated in the HRG model at $T = 155$~MeV and $\mu_B = 0$ with multiplicity-dependent magnetic field shown in Fig.~\ref{fig:eBLHC}.
    The ratios are normalized by their values in central collisions.
    }
    \label{fig:eBLHCNchratios}
\end{figure}

Measurements of the centrality dependence of hadron yield ratios other than $p/\pi$ can further test the possible magnetic field effect at chemical freeze-out.
The corresponding predictions of the HRG model with magnetized hadron spectrum are depicted in Fig.~\ref{fig:eBLHCNchratios}.
The upper panel depicts the $d/p$ and $n/p$ ratios, which becomes suppressed in the presence of magnetic field in non-central collisions.
This primarily stems, again, from the increased proton yield due to enhanced feeddown from $\Delta$.
The green band also shows experimental data of the ALICE Collaboration for $d/p$~\cite{ALICE:2022veq}, indicating a non-monotonic multiplicity dependence and possible suppression in central collisions. 
This qualitative behavior disfavors the magnetic field, although the errors are still large. 
It should be noted, however, that this analysis is based on the assumption of the thermal production mechanism for light nuclei~(see, e.g., ~\cite{Braun-Munzinger:2018hat} for an overview) and may not apply if another mechanism, such as coalescence, is assumed instead.

The middle panel of Fig.~\ref{fig:eBLHCNchratios} shows the multiplicity dependence of $K^+/\pi^+$~(solid red line) and $\phi/\pi^+$~(dash-dotted black line), along with the experimental data on $K/\pi$ in 5.02~TeV Pb-Pb collisions~\cite{ALICE:2019hno}.
The magnetic field has a weak effect on final yields of pions, kaons, and $\phi$. Thus, the normalized ratios stay close to unity in the HRG model calculation.
On the other hand, the experimental data on $K/\pi$ ratio show indications of mild suppression at low multiplicities.

The bottom panel of Fig.~\ref{fig:eBLHCNchratios} depicts the multiplicity dependence of various hyperon-to-pion ratios: $\Lambda/\pi^+$~(solid orange line), $\Xi^-/\pi^+$~(dashed blue line), and $\Omega^-/\pi^+$~(dash-dotted cyan line).
All three ratios are enhanced by the magnetic field, with the enhancement of the $S = 3/2$ $\Omega^-$ baryon being the most prominent.
The relevance of the magnetic field can thus be probed by the possible stronger enhancement of $S = 3/2$ baryons, such as $\Omega^-$ over that of $S = 1/2$ baryons.
A measurement of $\Delta(1232)^{++}/p$ ratio as a function of centrality could also be instructive, although the final yields of short-lived resonances such as $\Delta(1232)^{++}$ may be strongly affected by the hadronic phase~\cite{Motornenko:2019jha,Oliinychenko:2021enj}.

Perhaps most promising in this regard, from a theoretical perspective, would be the measurements of neutron-to-proton ratio, $n/p$, which exhibits a considerable suppression~($\sim 10-15$\%) in the presence of magnetic field.
This is a manifestation of the isospin symmetry breaking by the magnetic field.
Other centrality-dependent effects, such as baryon annihilation, are expected to maintain the isospin symmetry, thus, the $n/p$ ratio may be the most sensitive and unambiguous probe among the hadron yield ratios discussed here.
It should be noted that the suppression of $n/p$ ratio can be studied not only on the level of integrated yields, but also at a fixed transverse momentum $p_T$.
Although the modeling employed here is not sophisticated enough to provide $p_T$-distributions, based on the fact that the magnetic field mainly affects protons and neutrons through the feeddown from $\Delta(1232)$ decays, the strongest suppression can be expected at low $p_T$.
This analysis could motivate the corresponding experimental measurements of neutrons in the future, which are, of course, extremely challenging.

\section{Magnetic field and fluctuations}
\label{sec:fluctuations}

This section explores the effect of a magnetic field on conserved charge susceptibilities and measurable fluctuations of identified hadrons.
The results are discussed in the context of earlier works on the subject~\cite{Bhattacharyya:2015pra,Ding:2023bft}, and expand on them by considering variations on the hadron list and the presence of repulsive interactions.

\begin{figure}[t]
    \includegraphics[width=.49\textwidth]{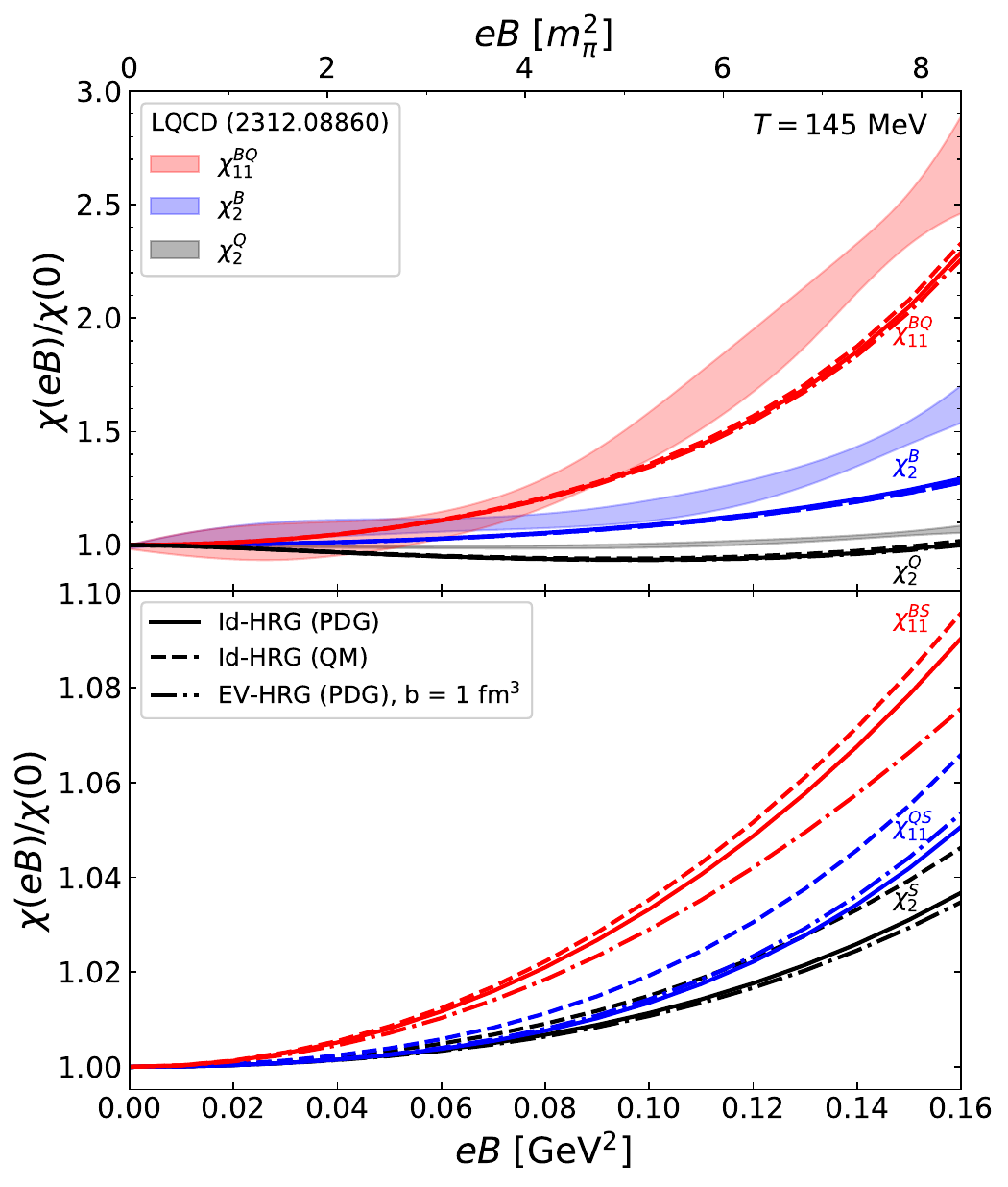}
    \caption{
    The dependence of second-order susceptibilities of conserved charges $\chi_2^B$, $\chi_{11}^{BQ}$, $\chi_2^Q$~(upper panel), $\chi_{11}^{QS}$, $\chi_{11}^{BS}$, and $\chi_2^S$~(lower panel) on the magnetic field along the temperature $T = 145$~MeV calculated within ideal HRG model with PDG2020~(solid lines) or quark model~(dashed lines) hadron list, as well as EV-HRG model (PDG2020 list) with $b = 1$~fm$^3$~(dash-dotted lines).
    The susceptibilities are normalized by their values at vanishing magnetic field, $eB = 0$.
    The bands depict lattice QCD data from Ref.~\cite{Ding:2023bft}.
    }
    \label{fig:chi2s}
\end{figure}

Figure~\ref{fig:chi2s} shows the magnetic field dependence of all second-order susceptibilities of conserved charges along the isotherm $T = 145$~MeV. 
The temperature value is taken to enable comparisons with recent lattice QCD continuum estimate from~\cite{Ding:2023bft}, shown in Fig.~\ref{fig:chi2s} by the bands.
The susceptibilities are normalized by their values for the vanishing magnetic field, $eB = 0$, i.e. $\hat{\chi}_{ij} = \chi_{ij}(eB) / \chi_{ij}(eB=0)$.
Calculations are performed within the ideal HRG model utilizing the default PDG2020 hadron list in Thermal-FIST~(solid lines) as well as the list containing extra states based on quark model~(dashed lines)~\cite{Alba:2017mqu,Alba:2020jir}. 
Extra states can have a notable effect on the absolute value of the susceptibilities and the resulting comparisons with lattice QCD~\cite{Bollweg:2021vqf}. However, these states' presence has a minor effect on their relative change due to the magnetic field, as Fig.~\ref{fig:chi2s} shows.
This is also the case for the HRG model incorporating a repulsive baryon core (EV-HRG)~\cite{Vovchenko:2017xad}.
Thus, the dependence of normalized second-order susceptibilities of conserved charges on the magnetic field appears to be robust with respect to the uncertainties in the hadron list and repulsive baryon interactions.

The HRG model reproduces the qualitative behavior seen in the lattice data, namely, $\hat{\chi}_{11}^{BQ} > \hat{\chi}_2^{B} > \hat{\chi}_2^Q$. In particular, the stronger enhancement of $\hat{\chi}_{11}^{BQ}$ compared to $\hat{\chi}_2^{B}$ can be attributed to the large enhancement of the $\Delta(1232)^{++}$ density in the HRG model, as previously discussed in~\cite{Ding:2023bft}. 
The HRG model, however, underpredicts the lattice data for $eB \gtrsim 0.08$~GeV$^2$.
This may indicate the need to improve the modeling of a magnetized HRG, as one would still expect the hadronic description to hold true at $T = 145$~MeV and not too strong magnetic fields. 
In particular, the present formulation of the model is questionable for hadrons with spin $S = 3/2$ or above, as the $S = 3/2$ hadrons would give a negative vacuum contribution to the pressure~\cite{Endrodi:2013cs}, marking instability of the theory.
One can also explore different values of gyromagnetic ratios for different hadrons.
Thus, a revised formulation of the model is required, which could modify the predictions for the susceptibilities of conserved charges.

\begin{figure}[t]
    \includegraphics[width=.49\textwidth]{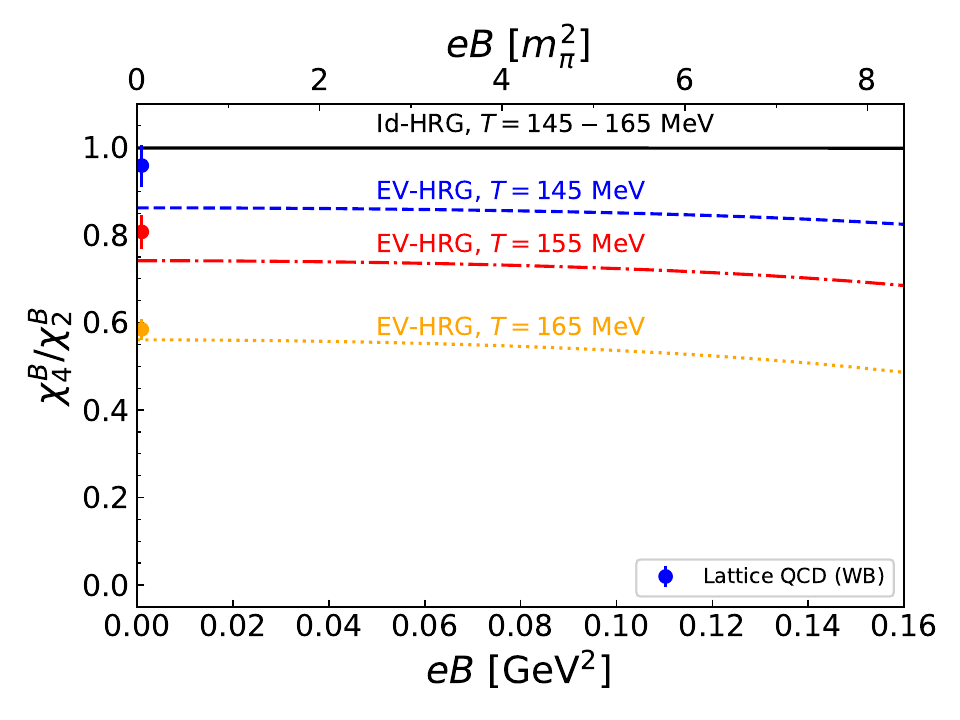}
    \caption{
    The dependence of fourth-to-second order susceptibility ratio $\chi_4^B/\chi_2^B$
    on external magnetic field $eB$ evaluated within ideal~(black line) and excluded volume~(colored lines) HRG model at $T = 145$, $155$, and $165$~MeV (top to bottom).
    Id-HRG model results are consistent with unity at all temperatures.
    The symbols show $N_t = 12$ lattice QCD results of the Wuppertal-Budapest collaboration for $\chi_4^B/\chi_2^B$ at $eB = 0$~\cite{Borsanyi:2018grb}.
    }
    \label{fig:chi4s}
\end{figure}

Figure~\ref{fig:chi4s} shows the behavior of the net baryon number kurtosis, $\chi_4^B / \chi_2^B$, as a function of magnetic field for three temperatures, $T = 145$, 155, and 165~MeV.
This ratio is consistent with unity in the ideal HRG model for both PDG2020 and QM hadron lists, indicating that fluctuations of baryon number are largely governed by the Skellam distribution (Poisson statistics) not only at vanishing magnetic field but also in the presence of a finite magnetic field.
This can be seen by looking into Eq.~\eqref{eq:pc} for the partial pressures of hadrons at finite $eB$.
If quantum statistics is negligible -- a very good approximation for baryons at $\mu_B = 0$ corresponding to keeping only the $k = 1$ term in Eq.~\eqref{eq:pc} -- the $\mu_B$ dependence of the pressure enters through $e^{\pm \mu_B/T}$ term only, as in the ideal HRG model for $eB = 0$.
The above observation implies that the growth of second-order baryon susceptibilities $\chi_2^B$, $\chi_{11}^{BQ}$, and $\chi_{11}^{BS}$ is driven entirely by the increase of hadron number densities in the ideal HRG model, and not by the appearance of any dynamical fluctuations.

In the EV-HRG model, baryon fluctuations are suppressed by the presence of a repulsive baryon core.
For $b = 1$~fm$^3$ the model can describe the systematics of lattice QCD data on diagonal baryon susceptibilities at temperatures $T \lesssim 165$~MeV for $eB = 0$~\cite{Karthein:2021cmb}~(see also the points in Fig.~\ref{fig:chi4s}).
In the presence of a magnetic field, $\chi_4^B/\chi_2^B$ in the EV-HRG model shows mild suppression for sufficiently large $eB \gtrsim 0.16$~GeV$^2$.
This suppression is attributed to the growth of the density of baryons and antibaryons with $eB$, which makes the excluded volume effect stronger.
Therefore, lattice QCD calculations of $\chi_4^B/\chi_2^B$ for the sufficiently large magnetic field can be used to probe the relevance of the repulsive baryon core.
It can be noted that all HRG model results for $\chi_4^B/\chi_2^B$ presented in Fig.~\ref{fig:chi4s} are identical for mixed susceptibility ratios $\chi_{31}^{BQ}/\chi_{11}^{BQ}$ and $\chi_{31}^{BS}/\chi_{11}^{BS}$.
It has been argued in Ref.~\cite{Karthein:2021cmb} that these additional ratios can be used to probe the flavor dependence of interactions in the HRG model.

In heavy-ion collisions it is challenging to measure fluctuations of conserved charges with the exception of net-charge number.
Instead, various proxies are used in place of conserved charges~\cite{Kitazawa:2012at,Bellwied:2019pxh}.
It has been argued that measuring fluctuations of various proxies in different centralities can probe the magnetic field in heavy-ion collisions~\cite{Ding:2021cwv,Ding:2023bft}.

Here the behavior of net-proton, net-kaon, and net-pion number fluctuations is explored. 
The corresponding measurements are being performed by the ALICE Collaboration~\cite{ALICE:2019nbs,ALICE:2022xpf}, focusing on the normalized variance,
$R_1[N_\pm] = \frac{\kappa_2[N_+ - N_-]}{\mean{N_+ - N_-}}$,
where $N_\pm$ are the numbers of (anti)particles.
The denominator is the normalization factor, such that the baseline $R_1[N_{\pm}] = 1$ corresponds to uncorrelated particle production.
At LHC energies, $R_1$ can be connected to the $\nu_{\rm dyn}$ measure~\cite{Pruneau:2019baa}.

\begin{figure}[t]
    \includegraphics[width=.49\textwidth]{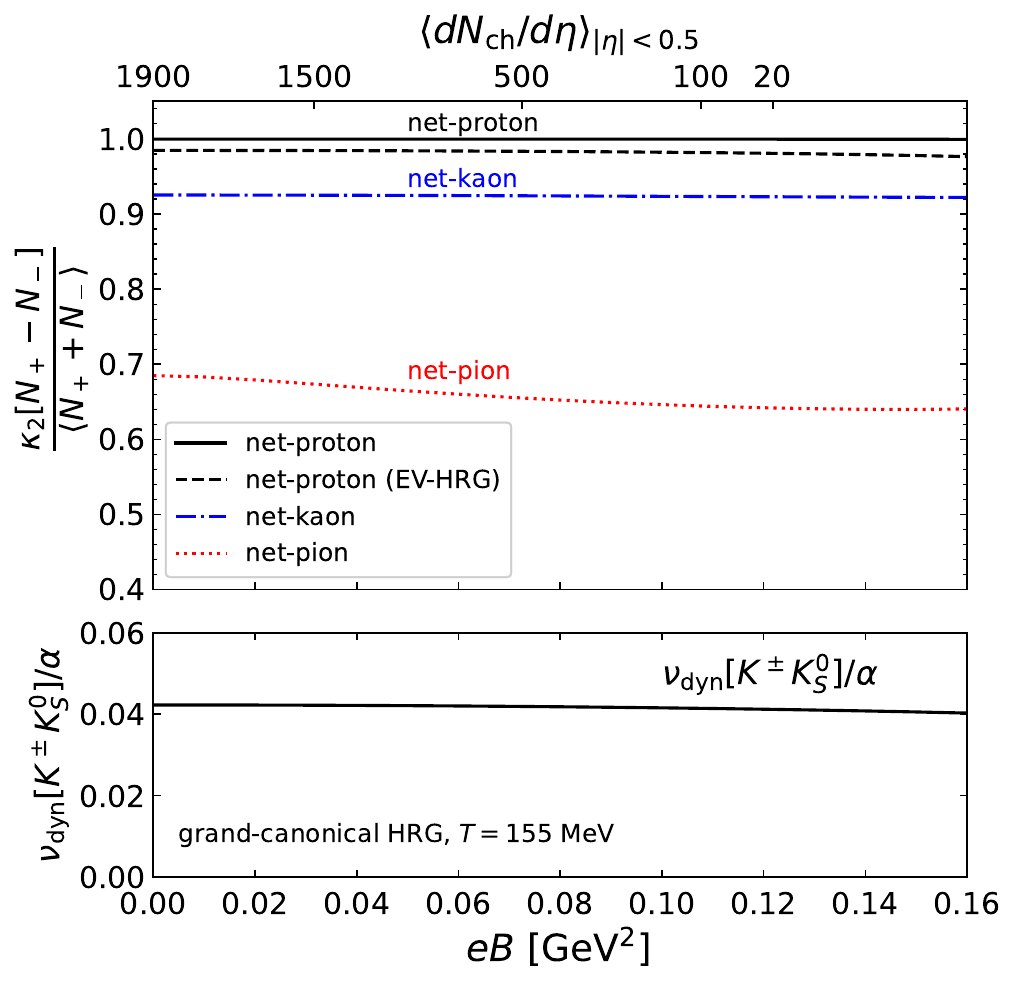}
    \caption{
    The upper panel shows the magnetic field dependence of normalized variance $\kappa_2[N_+ - N_-]/\mean{N_+ + N_-}$ of net-proton (solid black), net-kaon~(dash-dotted blue), and net-pion~(dotted red) numbers calculated within the ideal HRG model in the grand-canonical ensemble at $T = 155$~MeV.
    The black dashed line depicts the calculations for net-proton number within the EV-HRG model.
    The lower panel shows the quantity $\nu_{\rm dyn}[K^{\pm} K_S^0] / \alpha$.
    The upper horizontal axis shows the corresponding charged multiplicity in Pb-Pb collisions at $\sNN = 5.02$ TeV, assuming a multiplicity-dependent magnetic field as shown in Fig.~\ref{fig:eBLHC}.
    }
    \label{fig:chi2LHC}
\end{figure}

The upper panel of Fig.~\ref{fig:chi2LHC} depicts the normalized variance $\kappa_2[N_+ - N_-]/\mean{N_+ + N_-}$ of net-proton (solid black), net-kaon~(dash-dotted blue), and net-pion~(dotted red) numbers calculated within the ideal HRG model in the grand-canonical ensemble at $T = 155$~MeV. The black dashed line additionally depicts the calculations for net-proton number within the EV-HRG model.
One can see that the magnetic field has negligible dependence of normalized variance of net-proton and net-kaon number, and slightly suppressed the normalized variance of net-pion number.
Although the magnetic field does enhance the variance $\kappa_2[p-\bar{p}]$ of net-proton number itself, reflecting the enhancement of baryon susceptibilities shown in Fig.~\ref{fig:chi2s}, this enhancement is canceled out by the identical enhancement of the normalization factor $\mean{p + \bar{p}}$.
It can, therefore, be concluded that, in the HRG regime, the magnetic field does not introduce any additional dynamical correlations of protons, meaning that the corresponding measurements of fluctuations are unlikely to offer any significant information about the magnetic field in addition to that carried by mean hadron yields.
The normalized variances of net-kaon and net-pion numbers are considerably below unity due to correlations generated through resonance decays, as discussed in Refs.~\cite{Vovchenko:2020kwg,ALICE:2022xpf} for the case of zero magnetic field.

It should be noted that the above calculations are performed within the grand-canonical HRG model without any kinematic cuts, thus, they are not directly applicable for comparisons to the experiment.
Nevertheless, they demonstrate these observables' (in)sensitivity to the magnetic field.
The experimental measurements indicate mild suppression of the normalized net proton variance with respect to unity, which is largely interpreted as being driven by (local) baryon number conservation giving $\frac{\kappa_2[p - \bar{p}]}{\mean{p + \bar{p}}} = 1 - \alpha$ where $\alpha$ is the acceptance factor~\cite{ALICE:2019nbs,ALICE:2022xpf}.
The calculations indicate that the canonical ensemble baseline should similarly be described through the same $1-\alpha$ factor in the presence of the magnetic field, with the possible modification of the value of $\alpha$ in different momentum space acceptances.

The possible effect of the magnetic field on correlations among kaons is also explored here.
Earlier, the ALICE Collaboration reported anomalously large correlations between charged and neutral kaons as a function of centrality~\cite{ALICE:2021fpb}.
The lower panel of Fig.~\ref{fig:chi2LHC} depicts the behavior of normalized correlator $\nu_{\rm dyn}[K^{\pm}K_S^0]/\alpha$ [with  $\alpha = (\mean{K_S^0}^{-1} + \mean{K^\pm}^{-1})^{-1}$] in the grand-canonical HRG model as a function of magnetic field at $T = 155$~MeV.
This quantity is essentially flat as a function of $eB$, largely ruling out the magnetic field as a possible explanation of the data.

\begin{figure}[t]
    \includegraphics[width=.49\textwidth]{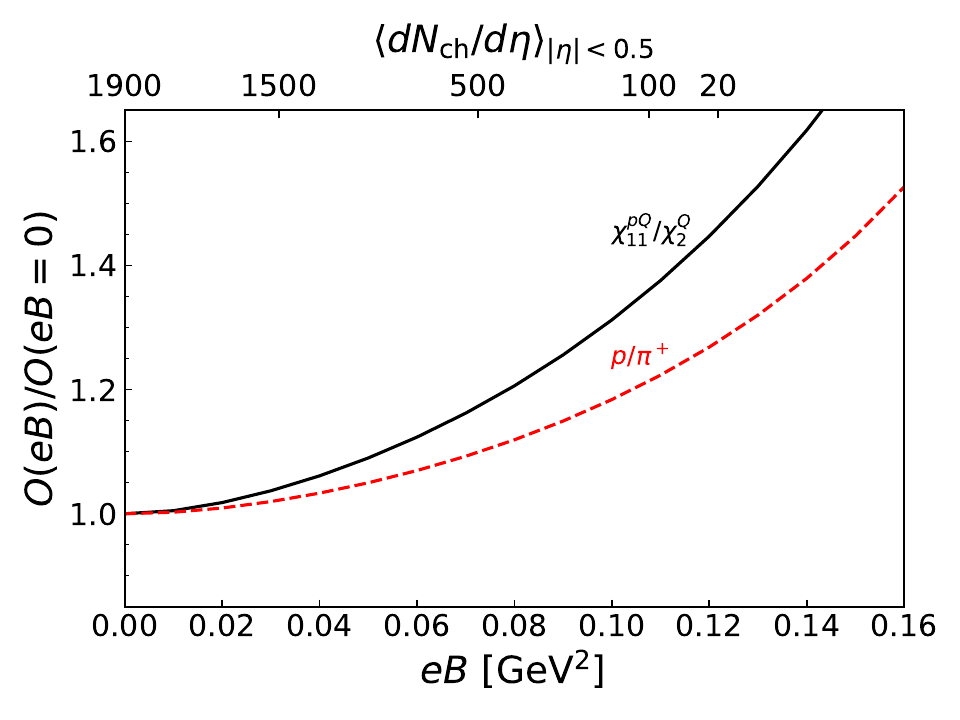}
    \caption{
    The dependence of susceptibility ratio $\chi_{11}^{pQ}/\chi_2^Q$ and the $p/\pi$ ratio on the magnetic field $eB$, normalized by the corresponding values at vanishing magnetic field $eB = 0$.
    The correlator $\chi_{11}^{pQ}$ includes resonance decay feeddown.
    The upper horizontal axis shows the corresponding charged multiplicity in Pb-Pb collisions at $\sNN = 5.02$ TeV, assuming a multiplicity-dependent magnetic field as shown in Fig.~\ref{fig:eBLHC}.
    Calculations are performed at $T = 155$~MeV.
    }
    \label{fig:HRGpQ}
\end{figure}

Finally, the normalized proton-charge correlator, $\chi_{11}^{pQ} / \chi_2^Q$, which is a proxy for baryon-charge correlator $\chi_{11}^{BQ} / \chi_2^Q$ is analyzed. This quantity has been suggested as a possible magnetometer in heavy-ion collisions~\cite{Ding:2023bft}.
Figure~\ref{fig:HRGpQ} depicts the dependence of the ratio $\chi_{11}^{pQ} / \chi_2^Q$ on the magnetic field, normalized by its value at vanishing magnetic field.
The ratio increases with the magnetic field, meaning that the enhancement in non-central collisions may indeed indicate the presence of the magnetic field.
Qualitatively, in the HRG model $\chi_{11}^{pQ} \sim \mean{p}$ due to proton self-correlation, while the net charge susceptibility is driven by pion multiplicity, $\chi_2^Q \sim \mean{\pi}$.
Indeed, one can see in Fig.~\ref{fig:HRGpQ} that the enhancement of $\chi_{11}^{pQ} / \chi_2^Q$ mirrors that of the $p/\pi$ ratio.
Therefore, qualitatively, one can expect the enhancement of $\chi_{11}^{pQ} / \chi_2^Q$ by any mechanism that leads to the enhancement of $p/\pi$ ratio, it does not necessarily has to be the magnetic field.
One can see that the relative enhancement of $\chi_{11}^{pQ} / \chi_2^Q$ is stronger than that of $p/\pi$ ratio at a given value of the magnetic field.
This can be attributed to the $p\pi^+$ correlation contribution to the proton-charge correlator which is coming from $\Delta^{++} \to p + \pi^+$ decays.
Thus, the relevance of the magnetic field could potentially be probed by simultaneous measurements of $\chi_{11}^{pQ} / \chi_2^Q$ and $p/\pi^+$ ratios and the analysis of the difference between the two.

\section{Discussion and summary}
\label{sec:summary}

This work explored the influence of magnetic field on hadron yields and fluctuations in the framework of the HRG model and possible consequences for heavy-ion collisions. Pertaining to the obtained results, one can highlight the following observations:

\begin{itemize}
    \item The presence of a considerable magnetic field~[$eB \sim O(m_\pi^2)$] in a hadron resonance gas leads to sizable modifications of hadron densities, especially the $\Delta(1232)^{++}$ resonance, which feeds down into the final proton yields. If such a magnetic field exists at the freeze-out stage of heavy-ion collisions, it will lead to notable changes in various hadron yield ratios at different centralities. In particular, the magnetic field leads to an enhancement of $p/\pi$ ratio in peripheral collisions, which is in line with the ALICE measurements performed for  5.02 TeV Pb-Pb collisions. By attributing the $p/\pi$ enhancement entirely to the magnetic field, an upper limit of $eB \simeq (6.6 \pm 0.6) m_\pi^2$ is obtained for peripheral collisions.
    Other hadron yield ratios can further probe this effect, in particular, the suppression of the $n/p$ ratio due to the breaking of isospin symmetry induced by the magnetic field.

    \item The increase in hadron densities at non-zero $eB$ drives the enhancement of susceptibilities of conserved charges, especially that of baryon-charge correlator $\chi_{11}^{BQ}$, which is in qualitative agreement with lattice QCD~\cite{Ding:2023bft}. However, the magnetic field does not induce any dynamical correlations in the HRG picture. Thus, the measurements of fluctuations and correlations in heavy-ion collisions appear to offer limited additional information about the magnetic field relative to that contained in mean multiplicities if the HRG picture applies. 
    The HRG model does show some deviations from lattice data~(Fig.~\ref{fig:chi2s}), which may potentially be probed experimentally.
    The enhancement of $\Delta(1232)^{++}$ yields due to magnetic field can be probed by simultaneous measurements of $\chi_{11}^{pQ} / \chi_2^Q$ and $p/\pi^+$ ratios and the analysis of the relative enhancement of the former with respect to the latter as a function of centrality.
    
\end{itemize}

The presented description can be improved and extended in a number of ways.
Although the present HRG model appears to reproduce the systematics of second-order susceptibilities in lattice QCD, the quantitative description at $T = 145$~MeV is far from perfect~(Fig.~\ref{fig:chi2s}).
The instability of the model with regard to $S = 3/2$~(and above) particles has been pointed out a long time ago~\cite{Endrodi:2013cs}, and an improved description is desirable.
For instance, the modifications of the masses of neutral hadrons, decay rates, and finite resonance widths could all play important roles.
In addition, one can consider variations in the gyromagnetic ratios of the different hadron species.
Furthermore, one has to explore the possible consequences that the pressure anisotropy in the presence of a magnetic field may have on heavy-ion observables.

Applications to heavy-ion collisions can be extended as well. 
In particular, one might regard the magnetic field as an additional fit parameter in thermal fits to hadron abundances produced at various collision energies and centralities, assuming that all other physical mechanisms are under control. 
This can readily be achieved through the open-source implementation in Thermal-FIST v1.5 discussed in Sec.~\ref{sec:HRG}.
One can also extend the modeling to account for large event-by-event fluctuations of the magnetic field and its inhomogeneous distribution along the freeze-out surface.
To enable studies of $p_T$-differential observables sensitive to the magnetic field, one can incorporate the magnetic field into a blast-wave description of hadron spectra at Cooper-Frye particlization.
In addition, a Monte Carlo sampler that takes into account the magnetic field effect can be developed and used for studies of fluctuations and correlations with realistic kinematic cuts and canonical ensemble effects.
These extensions will be the subject of future works.

Calculations presented in this work has been obtained using the open-source package \texttt{Thermal-FIST}~\cite{Vovchenko:2019pjl} version 1.5 which is publicly available at~\cite{FIST1p5}.

\vskip5pt

\emph{Acknowledgments.} 
The author thanks H.-T.~Ding, G.~Endr{\H o}di, V.~Koch, K.~Redlich, and H.~Stoecker for fruitful discussions.
The author also acknowledges the participants of EMMI Rapid Reaction Task Force ``Fluctuations and Correlations of Conserved Charges: Challenges and Perspectives'', Nov 6-10, 2023, GSI, Darmstadt, Germany, for discussions that motivated the studies presented in this work.


\emph{Note added.}
A preprint~\cite{Marczenko:2024kko} discussing similar issues appeared simultaneously with the present work.

\bibliography{main}

\end{document}